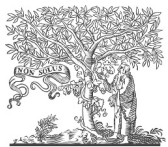
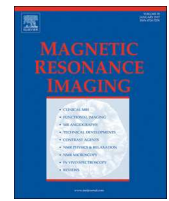

Original Contribution

# Demonstration of full tensor current density imaging using ultra-low field MRI

P. Hömmen*, J.-H. Storm, N. Höfner, R. Körber

*Physikalisch-Technische Bundesanstalt (PTB), Abbestraße 2-12, 10587 Berlin, Germany*

ABSTRACT

Direct imaging of impressed dc currents inside the head can provide valuable conductivity information, possibly improving electro-magnetic neuroimaging. Ultra-low field magnetic resonance imaging (ULF MRI) at µT Larmor fields can be utilized for current density imaging (CDI). Here, a measurable impact of the magnetic field $B_J$, generated by the impressed current density $J$, on the MR signal is probed using specialized sequences. In contrast to high-field MRI, the full tensor of $B_J$ can be derived without rotation of the subject in the scanner, due to a larger flexibility in the sequence design.

We present an ULF MRI setup based on a superconducting quantum interference device (SQUID), which is operating at a noise level of 380 aT Hz$^{-1/2}$ and capable of switching all imaging fields within a pulse sequence. Thereby, the system enables zero-field encoding, where the full tensor of $B_J$ is probed in the absence of other magnetic fields. 3D CDI is demonstrated on phantoms with different geometries carrying currents of approximately 2 mA corresponding to current densities between 0.45 and 8 A/m$^2$. By comparison to an *in vivo* acquired head image, we provide insights to necessary improvements in signal-to-noise ratio.

## 1. Introduction

*In vivo* current density imaging (CDI) of impressed currents in the mA range has a variety of possible applications, such as optimization and planning of therapeutic treatments like transcranial current stimulation with direct and alternating currents (tDCS/tACS) [1–6], or deep brain stimulation [7]. Further, three-dimensional conductivity maps, which are required for accurate source estimation in electro-magnetic neuroimaging [8], may be calculated from the current density distributions. In addition, conductivity imaging may have a value of its own, because it displays a physiologically meaningful physical tissue property.

Magnetic resonance imaging (MRI) can be used for non-invasive CDI [9–14]. Here, local magnetic fields $B_J$ generated by the impressed current density $J$ result in a measurable phase change of the MR signal. In high-field MRI at MHz Larmor frequencies, $B_J$ is approximately 7–9 orders of magnitudes smaller than the static $B_0$ field, limiting the detection of static magnetic fields to the component along the $B_0$ direction. To image all components of $B_J$, a rotation of the subject inside the scanner is necessary. Methods making use of a resonance phenomenon by time varying current stimulation at the Larmor frequency partly overcome this limitation [13,14]. However, conductivity values are obtained at frequencies far beyond physiological relevance. Another approach is the combination of MRI with surface potential measurements as performed in electrical impedance tomography (MR EIT) [15–17], which requires current injection from several directions. Conductivity tensors have been reconstructed *in vivo* using MR EIT in combination with diffusion tensor imaging [18].

Superconducting quantum interference device (SQUID)-based ultra-low field (ULF) MRI, working at µT fields (kHz Larmor frequencies) offers new possibilities for CDI [19–21]. The higher flexibility in sequence design due to the ability to switch all MRI fields within a pulse sequence enables full tensor $B_J$ imaging using a single unipolar impressed current and without subject rotation. However, the small magnetization in the µT regime poses special restrictions unique to ULF MRI. A pre-polarization pulse in the order of several tens of mT is essential to gain sufficient signal-to-noise ratio (SNR) [22]. Besides, the sensitive SQUID instrumentation requires shielding from the earth's magnetic field and environmental noise, in most cases achieved by mu-metal chambers and additional radio frequency shielding [23]. Regarding CDI in the human brain, the SNR is most crucial, because the strength of applied currents $J$ is limited by safety regulations to the low mA range [24,25]. Methods for CDI in the ultra-low field regime have been suggested for AC [26,27] and DC currents [27–29]. So far, the influence of impressed currents on 2D spin density and phase images has been presented for AC [26] and DC [27] currents, respectively.

---

* Corresponding author.
*E-mail address:* peter.hoemmen@ptb.de (P. Hömmen).

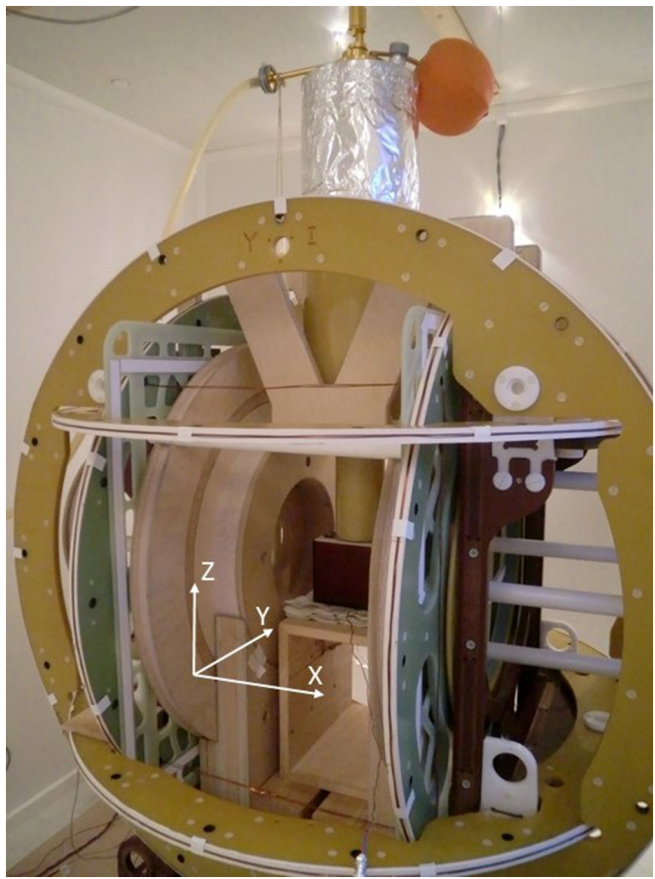

**Fig. 1.** The MRI setup for zero-field CDI consists of a 3-axis coil system and a one channel 2$^{nd}$-order SQUID gradiometer oriented in the z-direction. It comprises a self-shielded polarizing coil in the x-direction and Helmholtz coils for homogeneous fields in all Cartesian directions. Further, a Maxwell coil generates the frequency gradient $G_x$ and two saddle coils generate phase-encoding pulses $G_y$ and $G_z$. The phantom is positioned in the center of the coil setup.

However, full tensor $B_J$ mapping and current density reconstructions have not been demonstrated up until now, mainly due to hardware limitations.

In this work, we present an ultra-low field MRI setup that allows sequences necessary for 3D full tensor magnetic field mapping of DC fields in the order of a few nT. We performed phantom experiments with DC currents in the low mA range, yielding 3D $J$ reconstructions. This study is a proof of principle, setting the path towards *in vivo* CDI.

## 2. The experimental ULF MRI setup

### 2.1. Hardware setup

The MRI experiments were performed inside a custom-designed magnetically shielded room, based on the commercially available AK3b (Vacuumschmelze GmbH & Co. KG, Hanau, Germany). It consists of 2 layers of mu-metal, one eddy current shield, and it is situated inside a radio frequency (RF)-shielded room made of 2 mm thick zinc-plated steel panels.

The ULF MRI setup, shown in Fig. 1, comprises a one-channel SQUID sensor positioned in a 3-axis coil system, which was developed to generate homogeneous fields in all Cartesian directions, an essential part for the sequence described in Section 2.2. Further, a self-shielded polarizing coil with a field-current ratio of approximately 1 mT/A provides a homogeneous field in the x-direction, while reducing the magnetization of the mu-metal walls and thereby additional transient fields to a minimum.

The sensor system utilizes a current sensor SQUID with additional positive feedback [30]. On-chip within the input circuit, a current limiter consisting of 16 hysteretic SQUIDs is implemented. The SQUID is housed inside a niobium capsule to shield it from the high polarizing fields. The integrated input coil is connected to a wire-wound 2$^{nd}$-order axial gradiometer with 45 mm diameter and 120 mm overall baseline. The diameter was designed to obtain close to optimum SNR for source depths corresponding to the cortex [31]. The SQUID probe is operated inside a custom-built ultra-low-noise fiberglass dewar [32]. The distance between the lower loop of the gradiometer and the outside of the flat bottom dewar was measured to be 12.9 mm at 4.2 K.

The dewar has no extra RF-shield to avoid additional thermal noise. Consequently, all wires entering the mu-metal chamber are adequately shielded and grounded to protect the setup from RF-interferences. Further, all coils connected to power amplifiers are decoupled and grounded during read out. The $B_0$ field and the frequency gradient $dB_x/dx$ are driven by home-built low-noise current sources, ensuring stability below 80 ppm. With the MRI system in operation, we measured a noise level of approximately 380 aT Hz$^{-1/2}$ at the imaging frequency of about 1645 Hz.

### 2.2. Sequence design and experimental implementation

MR-based CDI is always based on imaging the magnetic field $B_J$, associated with the impressed current density $J$. If the full tensor of $B_J$ is acquired, $J$ can be determined from Ampere's law. For a non-magnetic material, such as biological tissue this becomes:

$$J = \frac{1}{\mu_0} \nabla \times B_J, \quad (1)$$

where $\mu_0$ is the permeability of free space. The CDI sequence exploited in this work is displayed in Fig. 2. It is based on zero-field encoding designed by Vesanen et al. [28] and is capable of full tensor $B_J$ imaging. Here, the term zero-field refers to the absence of all MRI fields, including the $B_0$ field. In an ideal case, the precession axis and frequency are set only by $B_J$ and can be expressed in matrix form of Bloch's equation:

$$\frac{d\boldsymbol{m}}{dt} = \gamma \boldsymbol{m} \times \boldsymbol{B}_J = \boldsymbol{A}\boldsymbol{m}, \quad (2)$$

where $\boldsymbol{m} = [m_x, m_y, m_z]^T$ is the magnetization, $\gamma$ the gyromagnetic ratio and

$$A = \gamma \begin{bmatrix} 0 & B_{J,z} & -B_{J,y} \\ -B_{J,z} & 0 & B_{J,x} \\ B_{J,y} & -B_{J,x} & 0 \end{bmatrix}. \quad (3)$$

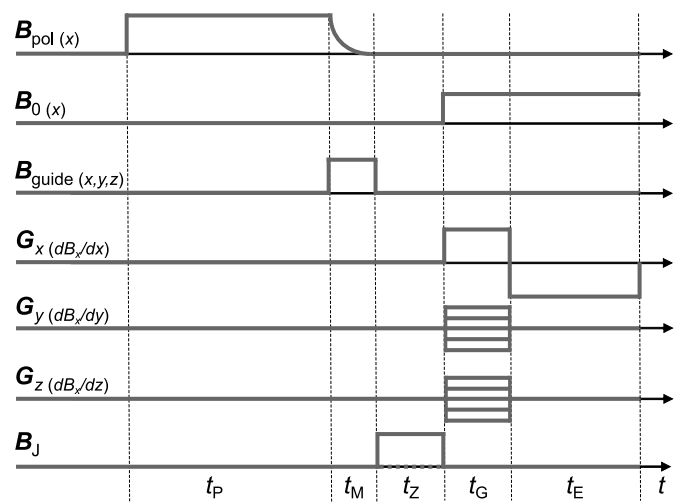

**Fig. 2.** Fourier gradient echo sequence for zero-field CDI. The sequence is divided into 5 stages ($t_P$ — polarization, $t_M$ — manipulating the orientation of the magnetization, $t_Z$ — zero-field encoding, $t_G$ — spatial encoding, and $t_E$ — echo time).

Introducing the time $t_Z$ during which $\boldsymbol{B}_J$ is present, a rotation matrix $\boldsymbol{\Phi} = e^{\boldsymbol{A} t_Z}$ can be used to describe the spin dynamics during the zero-field encoding period. Vesanen et al. give the details how $\boldsymbol{B}_J$ can be extracted from elements of $\boldsymbol{\Phi}$, which are derived from three different measurements with starting magnetizations $\boldsymbol{m}_0$ in the $x$-, $y$-, and $z$-direction, respectively. For example in the case of a starting magnetization in the $x$-direction, the relation becomes:

$$A = \begin{bmatrix} \Phi_{11} & \Phi_{12} & \Phi_{13} \\ \Phi_{21} & \Phi_{22} & \Phi_{23} \\ \Phi_{31} & \Phi_{32} & \Phi_{33} \end{bmatrix} \cdot \begin{bmatrix} |\boldsymbol{m}_0| \\ 0 \\ 0 \end{bmatrix} = [m_x, m_y, m_z]^{\mathrm{T}}. \quad (4)$$

Now assuming a sensor sensitive in the $z$-direction and $\boldsymbol{B}_0$ pointing in the $x$-direction, one can obtain $m_z$ and $m_y$ from the real and imaginary parts of the complex Fourier spectra, respectively, giving rise to

$$\Phi_{31} = \frac{S \cdot \cos(\rho - \delta)}{|\boldsymbol{m}_0|} \text{ and } \Phi_{21} = \frac{S \cdot \sin(\rho - \delta)}{|\boldsymbol{m}_0|}, \quad (5)$$

where $S$ denotes the voxel-specific magnitude values. The voxel-specific terms $\rho$ and $\delta$ describe the phase shifts due to $\boldsymbol{B}_J$ and the subsequent imaging sequence, respectively. The elements ($\Phi_{32}$, $\Phi_{22}$) and ($\Phi_{33}$, $\Phi_{23}$) can be extracted in the same manner, using the measurements with starting magnetizations in $y$- and $z$-directions.

In the ideal case discussed so far, no other fields during $t_Z$ are present and $\rho$ is solely determined by $\boldsymbol{B}_J$ and $t_Z$. The phase shift $\delta$ can be computed, at least in principle, from the parameters of the phase and frequency encoding.

However, the experimental situation is more complicated as other effects within the sequence influence $\rho$ and $\delta$. The unknown static and dynamic background fields over the sample volume during $t_Z$ will affect $\rho$. As these contributions cannot be assessed, one needs to perform a complete $\boldsymbol{B}$ reconstruction without applying $\boldsymbol{J}$, which can be subtracted subsequently.

The prediction of $\delta$ on the other hand suffers from inaccuracies in the imaging sequence such as non-ideal gradient ramps. The contributions to $\delta$ ought to be determinable by performing a measurement with $t_Z = 0$ s. However, in this case one cannot extract the true $\delta$ as for instance transient fields from the turn-off of the polarizing field extend now into the phase and frequency-encoding period. In Section 2.3, we show how we overcome this issue by using a calibration measurement.

A further requirement for successful CDI experiments is the avoidance of phase wrapping. The angle $\phi$ of the rotation $\boldsymbol{\Phi}$ is determined by the sum of $\boldsymbol{B}_J$ and the background field together with the duration of $t_Z$. For a given $\boldsymbol{B}_J$ and background field, phase wrapping can be prevented by choosing $t_Z$ accordingly ensuring $-\pi < \phi < \pi$.

The remaining steps in the extraction of $\boldsymbol{B}_J$ are performed as proposed by Ref. [28]. They comprise the calculation of the remaining row of $\boldsymbol{\Phi}$ by taking the cross product, conduct an orthogonalization to correct for noise, apply a non-linear inversion of the matrix exponential, and finally pick the corresponding elements in $\boldsymbol{A}$ to extract $\boldsymbol{B}_J$.

In summary, measuring $\boldsymbol{B}_J$ requires 6 runs of the full sequence illustrated in Fig. 2, since three different starting magnetizations before the zero-field period $t_Z$ are necessary and the later performed background field subtraction needs full imaging without applying $\boldsymbol{J}$. Furthermore, if the field mapping shall be assigned to the underlying structure, a seventh run is necessary to obtain an anatomical image unaffected by $\boldsymbol{B}_J$ and the background field. This can be achieved by setting $t_Z = 0$ s.

To accomplish the different magnetization orientations prior to the zero-field encoding, the polarizing field is turned off adiabatically into defined guiding fields $\boldsymbol{B}_{\mathrm{guide}}$ during the time $t_M$. In principle, the magnetization could also be manipulated by precisely designed tipping pulses as proposed by Vesanen et al. [28]. However, simulations have shown that our method is more stable against fluctuations in the shape and amplitude of the pulses. The adiabatic turn-off of the current in the polarizing coil is performed in two steps via a network of diodes and ohmic resistors. This results in a fast linear decrease at the beginning, turning into an exponential behavior once the induction voltage falls below the breakdown voltage of the diode. Fig. 3 illustrates the simulated spin dynamics during adiabatic turn-off of a 17 mT polarizing field in the $x$-direction into a 20 μT $\boldsymbol{B}_{\mathrm{guide}}$ in the $y$-direction. A custom-developed time domain Bloch equation solver based on the Runge-Kutta method was used for the calculations. One can see that for our setup a $\boldsymbol{B}_{\mathrm{guide}}$ of approximately 20 μT is sufficient for spin reorientation within $t_M = 15$ ms and an alignment error of approximately 0.5°. Note that the simulations incorporate relaxation $T_1 = T_2 = 100$ ms.

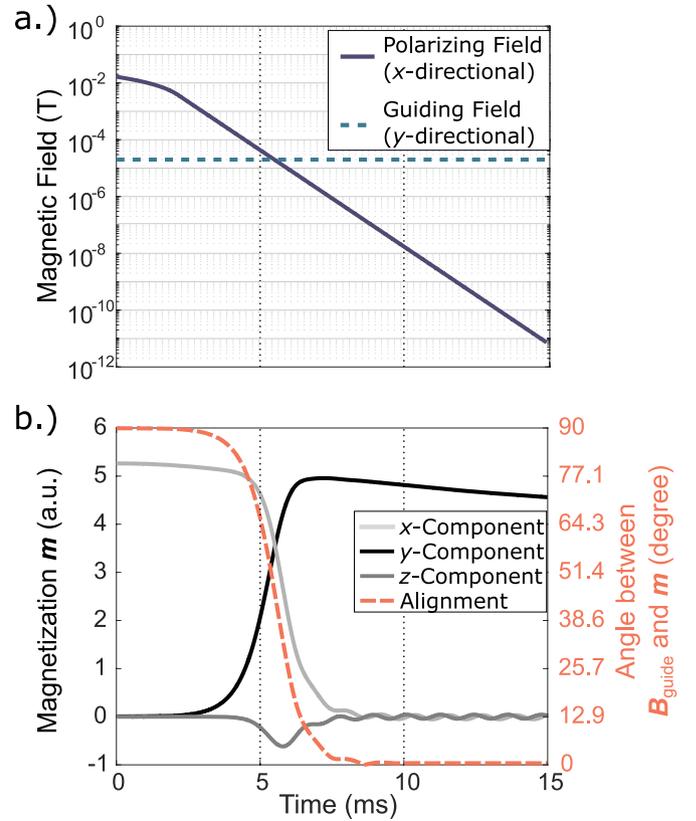

**Fig. 3.** Simulation of magnetization reorientation *via* adiabatic turn-off during $t_M$. a.) Time dependence of the simulated turn-off of a 17 mT polarizing field in the $x$-direction and a 20 μT guiding field in the $y$-direction. b.) Time dependence of the simulated evolution of the magnetization (left $y$-axis). The alignment between $\boldsymbol{m}$ and the guiding field is displayed in dashed red (right $y$-axis). (For interpretation of the references to color in this figure legend, the reader is referred to the web version of this article.)

Enlarging the guiding fields would reduce the alignment error. However, they could induce magnetization and eddy currents in the mu-metal walls yielding transient decays after turn-off, which are in parallel to $\boldsymbol{B}_{\mathrm{guide}}$. Therefore, setting the strength of $\boldsymbol{B}_{\mathrm{guide}}$ is a compromise between an alignment error of the starting magnetization and influences of the transient fields on spin dynamics during $t_Z$. Using a 3-axis fluxgate with a 3 dB bandwidth of 3 kHz (Mag-03MS, Bartington), we recorded the transient fields during $t_Z$ after ramping down $\boldsymbol{B}_{\mathrm{guide}}$. Fig. 4(a) illustrates the measured transient, consisting of a fast decay due to the current ramp down in the coil within 1 ms and the slower mu-metal response decreasing from about 90 nT down to < 1 nT in 15 ms. An approximation of the field transient by several combined exponential decays was used to simulate the influence on the evolution of the magnetization in the presence of a perpendicular $\boldsymbol{B}_J$ varying

Subsequent to the zero-field time, the $B_0$ field, the frequency gradient $G_x$ ($dB_x/dx$), and the phase-encoding gradients $G_y$ ($dB_x/dy$) and $G_z$ ($dB_x/dz$) are turned on for spatial encoding during $t_G$. Commercial power amplifiers switch the currents in the phase gradient coils within 1 ms. However, the $B_0$ field and $G_x$ are driven by low noise current sources with heavily filtered reference voltages, because they are active during signal acquisition. To ensure fast switching, we designed a dummy circuit mimicking the coils ohmic resistance, where the current is switched between coil and dummy load. The feedback circuit of the current source in combination with the $B_0$-coil has a 3 dB small-signal bandwidth of approximately 10 kHz. We achieve turn-off and recovery times of the 800 mA current in well below 1 ms. The stability of the current during read out was determined to about 80 ppm, measured via a monitor resistor.

The last step of the sequence is the read out of a gradient echo signal during the echo time $t_E$, generated by inverting $G_x$. In this time, the polarization coil, the guiding field coils, and the phase gradient coils are decoupled from the power amplifiers and grounded.

### 2.3. Phase calibration

The concept of the reconstruction of $J$ is based on the detection of measurable phase and amplitude changes in the MR signal due to $B_J$ and the background field. This requires complete knowledge of the system- and sample-specific MRI phase $\delta$, including all non-ideality in the imaging fields. As mentioned before, $\delta$ could be measured in principle by setting $t_Z = 0$ s. However, the change in the timing yields different influences by transient fields originating from the polarization. Instead, we chose a calibration method using a homogeneous magnetic field instead of $B_J$, in order to obtain the true $\delta$. Therefore, we generated 15 nT fields $B_{cal}$ using large Helmholtz coils and performed complete 3D measurements to reconstruct $B_{cal}$ and the background field. We chose a measurement without applying $B_{cal}$ to extract the voxel-specific MRI phase, which now was still influenced by the background field. Assuming the background field to be homogeneous over the sample volume, each voxel-specific $\delta$ was adjusted by a global offset phase until the reconstruction of $B_{cal}$ resulted in the correct orientation and amplitude. We found that this offset was purely system dependent, equal for $B_{cal}$ in either $x$-, $y$- or $z$-direction, and appeared stable over time. However, any changes in the coil system, timings, or field strengths will require a new calibration.

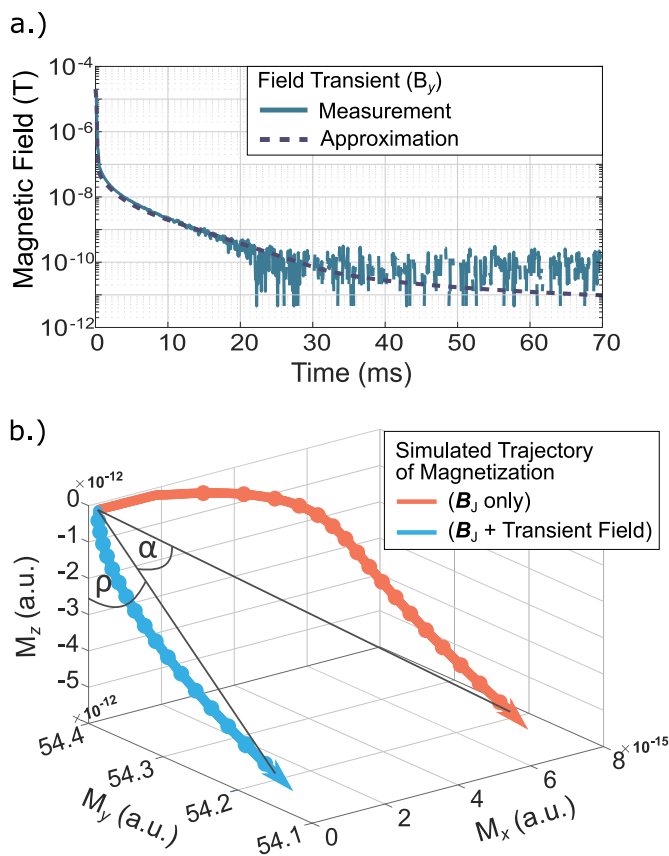

**Fig. 4.** The influence of transients generated by the guiding fields on spin dynamics during $t_Z = 70$ ms. a.) Fluxgate measurement of transient magnetic field in the $y$-direction after turn-off of $B_{guide,y}$ and approximation by combination of several exponential decays. b.) Simulated evolution of magnetization during the zero-field encoding period for $B_J$ only (light blue), and $B_J$ plus the transient field perpendicular to $B_J$ (red). The time scale is given by bold dots in 4 ms steps. The angles ρ and α represent the evolution of magnetization due to $B_J$ and the error angle due to the transient field, respectively. α was determined < 0.13 % of ρ, for $B_J$ between 1 and 50 nT. Please note the $x$-axis is blown up by three orders of magnitude. (For interpretation of the references to color in this figure legend, the reader is referred to the web version of this article.)

between 1 and 50 nT. An illustration for the case of $B_J = 10$ nT is given in Fig. 4(b). The results show that the transient in parallel to the starting magnetization influences spin dynamics only slightly, yielding an error angle smaller than 0.13 %. For reasons of illustration, these simulations were performed ignoring relaxation.

## 3. Measurements

### 3.1. $B_J$ mapping and current density reconstruction

3D CDI measurements were performed on phantoms with different geometries (Fig. 5 and Fig. 6(a)) utilizing the sequence described in

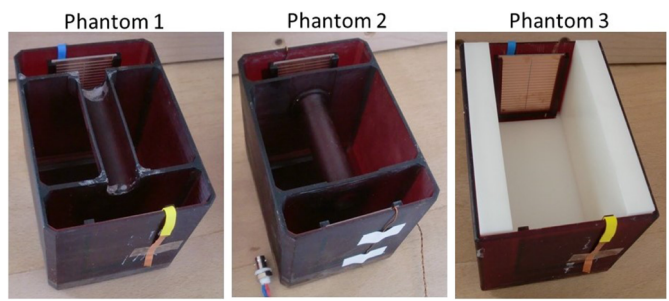

**Fig. 5.** Pictures of the phantoms used for the experimental validation of CDI. All phantoms have supports for (50 × 70) mm² electrodes. The Phantoms 1 and 2 possess a current path at the top and at a depth of 30 mm, respectively. Phantom 3 is a container allowing current flow through the entire volume.

Section 2.2. They were filled with an aqueous solution of $CuSO_4$ (0.079 wt%) tuning the relaxation time $T_1 = T_2$ to approximately 100 ms, a value similar to relaxation times of brain matter in the $\mu T$ regime [33,34]. Each phantom has supports for flat electrodes of the dimension $(50 \times 70)$ mm$^2$, which were fabricated from copper coated fiber reinforced plastic (35 $\mu$m copper). The leads were positioned such that their emitted magnetic field is lowest in the region below the sensor. Phantom 1 has the dimensions $(140 \times 100 \times 110)$ mm$^3$ and contains a small current pathway with a cross-sectional area of approximately 340 mm$^2$, yielding a nominal current density of 2.9 A/m$^2$ for 1 mA applied current. Phantom 2 with the same dimensions possesses a similar pathway (cross-sectional area of 315 mm$^2$, nominal current density 3.2 A/m$^2$ per mA applied current) 30 mm below the surface. Phantom 3 is a container with the dimensions $(140 \times 64 \times 70)$ mm$^3$, allowing current flow through the entire volume (nominal current density 0.22 A/m$^2$ per mA applied current).

We used a battery-powered, voltage-controlled current source to apply a 2.5 mA current to Phantom 1 and Phantom 2, resulting in mean current densities in the current paths of 7.4 A/m$^2$ and 7.9 A/m$^2$, respectively. The phantoms were positioned centrally in the MRI coordinate system (Fig. 1) such that the current flows in the negative $y$-direction in case of Phantom 1 and in the positive $x$-direction in case of Phantom 2. The distance from the top level of the solution to the bottom of the dewar was less than 1 cm, yielding a distance phantom-gradiometer of approximately 2 cm. CDI measurements were performed using 29 $k$-steps for each $y$- and $z$-direction, yielding a voxel size of $(4.8 \times 4.8 \times 4.8)$ mm$^3$ and a field of view FOV$_{y,z}$ = 139 mm. The strength of $B_0$ was set to 38.64 $\mu$T (1645 Hz). The frequency gradient $G_x$ was set to 121 $\mu$T/m and the maximum phase-encoding gradients $G_y$ and $G_z$ to $\pm$ 75.6 $\mu$T/m with a phase-encoding time $t_G$ of 30 ms. The current was applied to the phantom in the zero-field encoding time $t_Z$ = 70 ms. The polarization field was 17 mT and the polarizing time $t_P$ 500 ms including a 150 ms ramp up. As mentioned before, a full 3D $B_J$ mapping, together with a spin density image requires seven runs of the sequence per $k$-step, resulting in 5887 runs for 29 $\times$ 29 $k$-steps. With our current polarizing coil, we can perform approximately 1500 shots (17 mT) using a duty cycle of 1/3. Afterward, a cool down period of at least 2 h is necessary. Consequently, the measurement time in these experiments was approximately 9 h.

For the examination of physiologically realistic current densities, we applied a 2 mA current to Phantom 3, resulting in a mean current density of approximately 0.45 A/m$^2$. As voxel sizes of $(5 \times 5 \times 5)$ mm$^3$ did not result in sufficient SNR required for reliable $B_J$ reconstruction, we applied 15 $k$-steps for each $y$- and $z$-direction, yielding a voxel size of $(9.4 \times 9.4 \times 9.4)$ mm$^3$ in a field of view FOV$_{y,z}$ = 141 mm. The total of 1557 sequence runs was performed in about 55 min. The frequency gradient $G_x$ was set to 73 $\mu$T/m and the maximum phase-encoding gradients $G_y$ and $G_z$ to $\pm$ 36.5 $\mu$T/m with a phase-encoding time $t_G$ of 30 ms. The current was applied to the phantom in the zero-field encoding time $t_Z$ = 70 ms. The polarization and the $B_0$ field were equal to the previous experiments. The phantom was positioned in the MRI system to allow current flow in the negative $x$-direction.

### 3.2. Anatomical imaging

An *in vivo* implementation of the method requires both, high SNR and high resolution to resolve rather small structures such as scalp and skull. In order to compare the accessible SNR to the achievements in the phantom studies a 3D MR amplitude image of the human head was acquired using the CDI sequence with $t_Z$ = 0 s. Equal to the phantom measurements, the strength of the polarization field was 17 mT and the $B_0$ field was set to 38.64 $\mu$T (1645 Hz). The gradient fields were increased to achieve the maximum possible isotropic resolution with the current setup. The frequency gradient $G_x$ was set to 125 $\mu$T/m and the maximum phase-encoding gradients $G_y$ and $G_z$ to $\pm$ 95 $\mu$T/m with a phase-encoding time of 30 ms. We performed 35 $k$-steps for each $y$- and $z$-direction, resulting in a voxel volume of $(4.1 \times 3.9 \times 3.9)$ mm$^3$. The total measurement time was around 40 min.

## 4. Results and discussion

The results of the experiments with Phantom 1 and Phantom 2 are displayed in Fig. 6. At the top, a schematic shows the positioning in the MRI coordinate system (a). The second row (b) gives the MR amplitude images, obtained from the sequence with $t_Z$ = 0 s, in two different slices at the level of the current channels. The amplitudes are uncorrected for the sensitivity profile of the MRI system, which is dominated by the sensor sensitivity. The slice showing the $XY$ profile at $Z$ = 19.3 mm reveals two spots with low sensitivity that appear centrally on the $y$-axis, which originate from the sensitivity profile of the sensor in combination with a $B_0$ field in the $x$-direction. The outlines of the phantoms are visible in the images and drawn into the pictures as white dotted lines. In Fig. 6(c), the current density reconstructions $J$ are displayed for the same slices as the adjacent amplitude images, separated in $J_x$-, $J_y$-, and $J_z$-components. They were reconstructed from the imaged $B_J$ by utilization of Ampere's law (Eq. 1).

The experiment with Phantom 1 yielded a reconstruction of the current density $J$ in the negative $y$-direction. The amplitudes agreed well with the estimated mean current density of 7.4 A/m$^2$. The reconstructed current inside the current channel appeared distinguishable to the outside where reconstructions vary stochastically close to zero. A comparison to the MR amplitude images showed the reconstruction quality decreased with signal strength. The gray areas in the $J$-reconstruction images represent voxels where the MR amplitude fell below an empirically defined threshold of 1 fT. Because of the large variations in $B_J$ and thereby $J$ due to insufficient SNR, these areas were excluded for clarity reasons.

Fig. 6 also displays the results of the experiment with Phantom 2, where the current channel was 30 mm farther away from the sensor. Here, the phantom was positioned such, that the current inside the channel was $x$-directional. The outline of the current channel is observable in the MR amplitude images, albeit with significantly lower contrast than in the case of Phantom 1. This is due to the sensitivity, which decreases with distance to the sensor. The magnitudes of the reconstructed $J$ were in agreement with the predicted value of 7.9 A/m$^2$. Similar to the experiment with Phantom 1, the reconstructed current density was solely $x$-directional inside the current channel and distinguishable to the outside. However, we observed a larger variation of the current amplitudes, due to the reduced SNR.

Fig. 7 displays the results for MR amplitude (a) and $B_J$ reconstruction (b) in a $YZ$-slice centrally through Phantom 3. The phantom outline is indicated by dotted lines. In comparison to the previous experiments, voxel volumes were chosen about 7.5 times larger to gain sufficient SNR for the reconstruction of the 20 times lower current density. Maximum MR amplitudes were increased only by a factor of 3, caused by the non-uniform sensitivity profile of the sensor. The magnetic field distribution revealed a circulation in the $YZ$-plane, which is due to the applied current in the negative $x$-direction. However, the center of the circulation was not aligned with the center of the phantom, even though a nearly homogeneous current flow was assumed due to the phantom geometry. In simulations, not shown here, we found that this was due to the influence of the leads connecting the electrodes. These were combined to a twisted pair 2 cm below the phantom. The current density reconstruction, displayed in Fig. 7(c), was not affected by the field of the leads because the curl operator in Eq. 1 is insensitive to currents external to the phantom. Similar to the previous experiment, an empirically defined threshold of 10 fT in the MR amplitude image was used. In areas with sufficient SNR, $J$ appeared in the negative $x$-direction with magnitudes between 0.4 and 0.5 A/m$^2$, which is in agreement with the expected value calculated from the cross-sectional area. However, variations were larger than in the experiment with Phantoms 1 and 2.

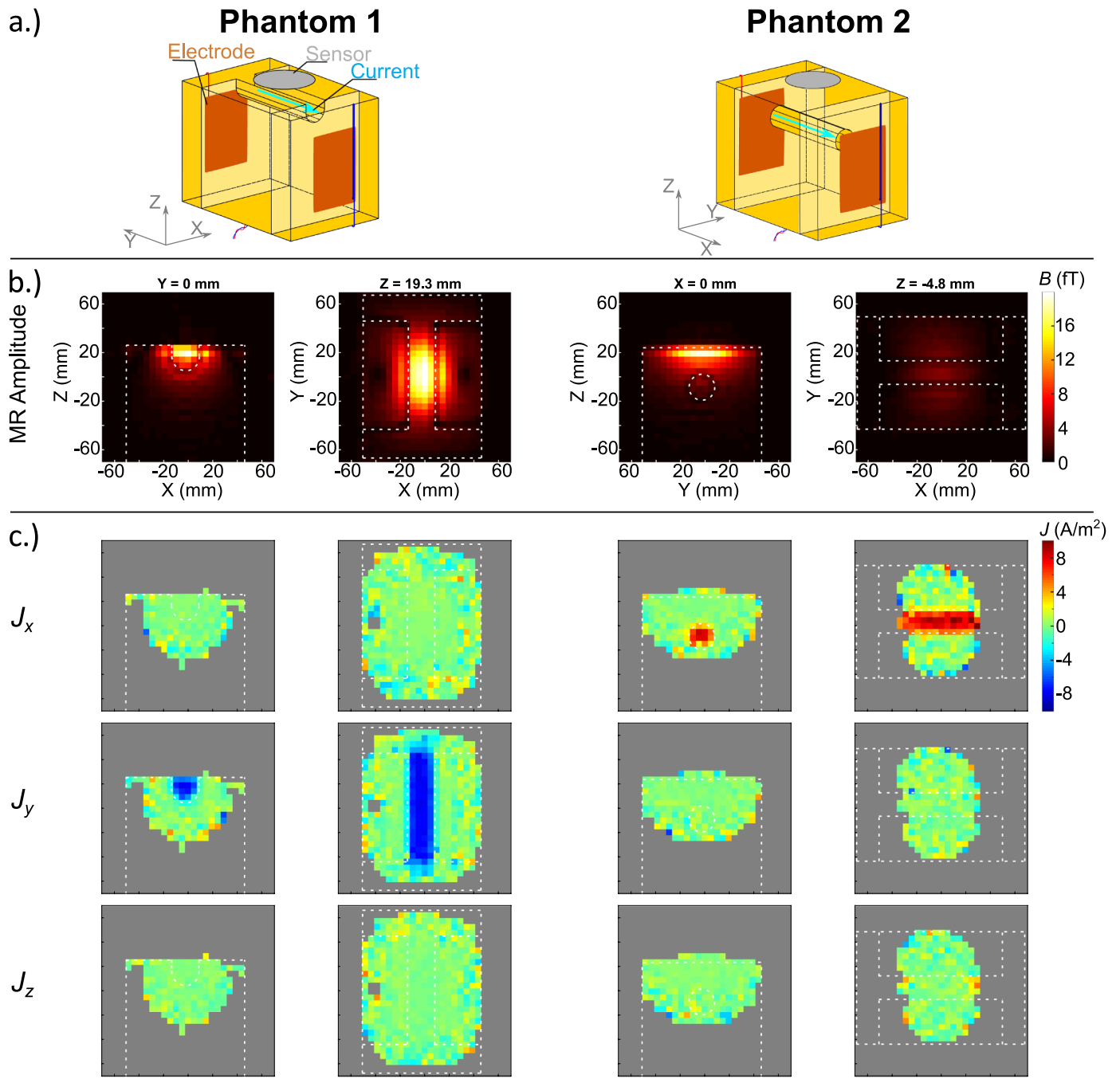

**Fig. 6.** 3D CDI on Phantom 1 and Phantom 2. a.) Schematics showing the positioning inside the MRI scanner. Both phantoms are placed centrally below the sensor, but in different orientations (see coordinate systems). The distance of the phantom to the lowest loop of the gradiometer was approximately 20 mm. b.) MR amplitudes in two different slices of each phantom, which are intersecting in the center of the current channels. The outlines of the phantoms are indicated by white dotted lines. c.) Current density reconstructions for $J_x$-, $J_y$-, and $J_z$-components respectively. The gray areas represent voxels that were excluded from the reconstruction because their corresponding MR amplitude was below a 1 fT threshold.

Fig. 8 presents the results of *in vivo* anatomical imaging. The sensor arrangement above the temporal lobe of the subject (a) is displayed along with the spin density reconstruction (b) uncorrected for the sensitivity profile of the sensor. Signal above noise level could be observed down to depths of approximately 5 cm below the dewar surface and about ± 8 cm in the *x*- and *y*-directions, depending on the depth. Anatomical structures like scalp, skull, and intracranial tissue could be distinguished.

The achieved voxel volume of (4.1 × 3.9 × 3.9) mm³ was slightly smaller than 3D ULF MR images of the human head reported so far [21,34,35]. A comparison with the CDI experiment using Phantom 3 reveals that the maximum voxel amplitudes were approximately ten times lower in case of the *in vivo* image. This was most probably due to a combination of smaller voxel sizes, partial volume effects, and spin density as well as $T_1$ differences between different tissue types and the phantom.

An improvement in SNR by a factor of 3 could be gained by enlarging the voxel size to approximately (6 × 6 × 6) mm³. Depending on the field distribution and the tissue structure, even larger volumes are possible using non-isotropic voxels. However, robust predictions require simulations with realistic field estimates as present in the head when injecting currents, which have to be addressed in future work. Nevertheless, a further boost in SNR is crucial for *in vivo* implementation of CDI.

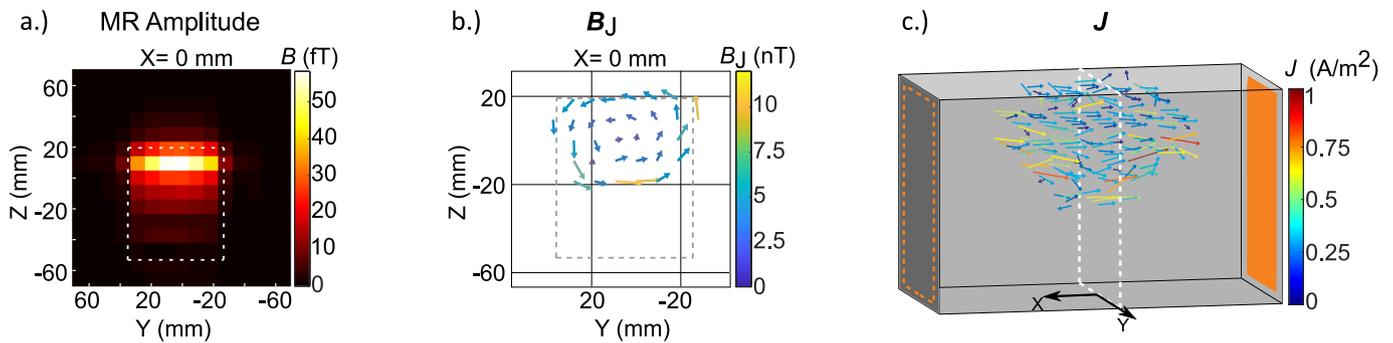

**Fig. 7.** 3D CDI on Phantom 3. a.) MR amplitudes, displayed in a *YZ*-slice. The outline of the phantom is indicated by white dotted lines. b.) $B_J$ reconstruction, where voxels with MR amplitude below 10 fT were excluded. c.) $J$ reconstruction in an overview plot of the entire phantom. The electrode dimensions are simplified by copper colored planes. The slice illustrated in (a) and (b) is indicated by white dotted lines. (For interpretation of the references to color in this figure legend, the reader is referred to the web version of this article.)

Considered that the noise floor of our system is already very low, substantial improvements in SNR are most likely possible by larger polarizing fields. Our polarizing system comprises a self-shielded room temperature coil in quasi Helmholtz configuration, similar to [36]. While this arrangement has multiple upsides in terms of field homogeneity and negligible transient fields after pulsing, it limits the field strength due to a higher resistance and the maximum number of applicable pulses due to overheating. Consequently, a cooled polarization coil with a high field-per-current ratio and acceptable homogeneity is required. Espy et al. [35] have reached 100 mT polarization fields using a liquid nitrogen cooled coil setup. Inglis et al. [37] reached field strengths of 80–150 mT over the head volume using a water cooled hollow copper tubing solenoid. Both systems described in Refs. [35,37] operated at Larmor frequencies > 5 kHz, due to excess low-frequency noise. Vesanen et al. [21] used a superconducting polarization coil able to produce 50 mT and possibly more, which however suffers from flux trapping in the superconducting filaments above 22 mT.

Besides improvements in SNR, applications such as *in vivo* conductivity mapping or optimization of tDCS would benefit much of a larger sample coverage. Multi-channel ULF MRI systems [34,35], or even whole-head sensor configurations for hybrid MEG-MRI [21] exist, albeit with inferior noise performance. The methods we applied to reduce noise interferences are transferable to multi-channel configurations.

## 5. Conclusion

In this study, we demonstrated that full tensor **B**-field mapping and thereby current density imaging is possible using MRI techniques in the ultra-low field regime. We could resolve current densities as low as 0.45 A/m$^2$ for a voxel size of $(9.4 \times 9.4 \times 9.4)$ mm$^3$ and 7.4 A/m$^2$ for a voxel size of $(4.8 \times 4.8 \times 4.8)$ mm$^3$. Key ingredients to the achievements are the ultra-low-noise performance of our ULF MRI system and the ability to switch all imaging fields including the $B_0$ field within a pulse sequence.

An *in vivo* implementation of the method will require further improvements regarding SNR and measurement time. To accomplish this, we have suggested to modify the polarization scheme. In addition, a larger sample coverage would be beneficial for many applications such as *in vivo* conductivity mapping and tDCS optimization.

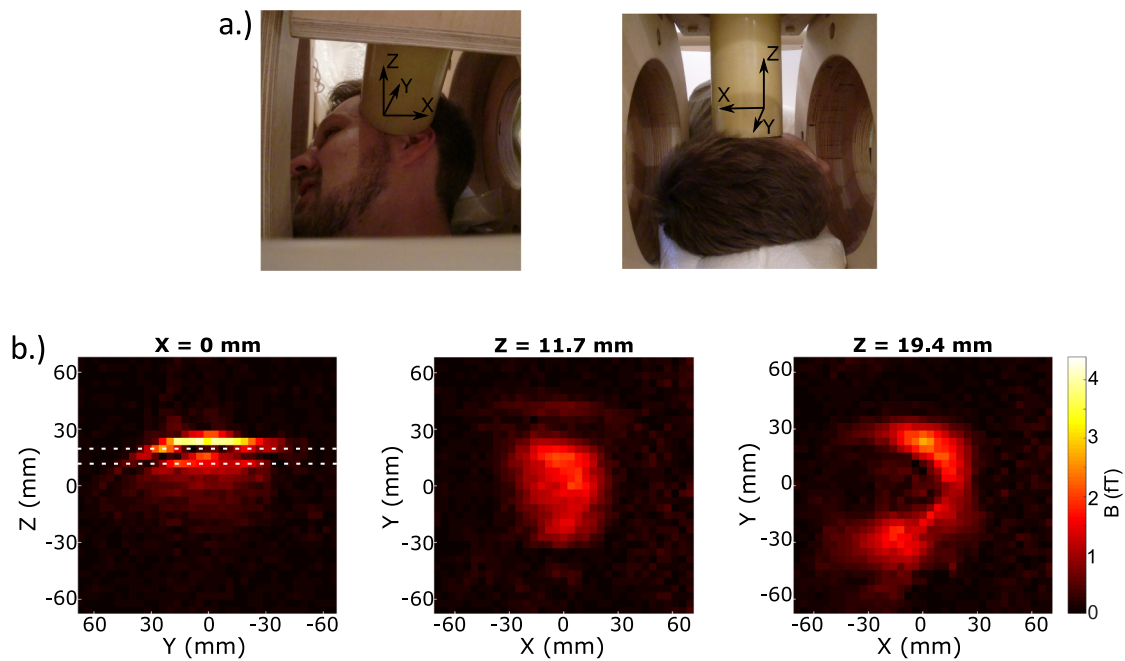

**Fig. 8.** Anatomical image of the human head. a.) Positioning of the subject. b.) MR amplitude image in the central *YZ*-slice (left), and in two *XY*-slices (middle and right).

**Acknowledgements**

This project has received funding from the European Union's Horizon 2020 research and innovation programme under grant agreement No 686865, and in part by the DFG under Grant KO 5321/1-1.